\documentclass[preprint,showpacs,aps,amsmath,amssymb]{revtex4-1}
\usepackage{epsf}
\usepackage{color}
\usepackage{dcolumn}
\usepackage{bm}
\everymath{\displaystyle}

\begin{document} 

\title{General properties of the Foldy-Wouthuysen transformation and applicability of the corrected original Foldy-Wouthuysen method}

\author{Alexander J. Silenko}
\email{alsilenko@mail.ru} \affiliation{Bogoliubov Laboratory of Theoretical Physics, Joint Institute for Nuclear Research, Dubna 141980, Russia,\\
Research Institute for Nuclear Problems, Belarusian State University, Minsk 220030, Belarus}


\begin {abstract}
General properties of the Foldy-Wouthuysen transformation which is
widely used in quantum mechanics and quantum chemistry are
considered. Merits and demerits of the original Foldy-Wouthuysen
transformation method are analyzed. While this method does not
satisfy the Eriksen condition of the Foldy-Wouthuysen
transformation, it can be corrected with the use of the
Baker-Campbell-Hausdorff formula. We show a possibility of such a
correction and propose an appropriate algorithm
of calculations. An applicability of the corrected
Foldy-Wouthuysen method is restricted by the condition of
convergence of a series of relativistic corrections.
\end{abstract}
 \maketitle

\section{Introduction}\label{Introduction}

The Foldy-Wouthuysen (FW) transformation first proposed in the seminal work \cite{FW} is now widely used not only in physics but also in quantum chemistry.
The FW representation has unique properties. In this representation, the Hamiltonian and all operators are even, i.e.,
block-diagonal (diagonal in two spinors). Relations between
the operators in the FW representation are similar to those
between the respective classical quantities. The form of quantum-mechanical operators for relativistic particles
in external fields is the same as in the nonrelativistic quantum theory. In particular, the position (Newton-Wigner)
operator \cite{NW} and the momentum one are equal to $\bm r$ and $\bm p=-i\hbar\nabla$, respectively. The polarization operator for
spin-1/2 particles is defined by the Dirac matrix $\bm \Pi$ and is expressed by much more cumbersome
formulas in other representations (see \cite{FW,JMP}). A great advantage of the FW representation is the simple
form of operators corresponding to classical observables.
The passage to the classical limit usually reduces to a replacement of the operators in quantum-mechanical Hamiltonians and equations of motion
with the corresponding classical quantities. The possibility of such a replacement, explicitly
or implicitly used in practically all works devoted to the FW transformation, was
rigorously proved in Ref. \cite{JINRLett12}. Thanks to these
properties, the FW representation provides
the best possibility of obtaining a meaningful classical limit
of relativistic quantum mechanics \cite{FW,CMcK}.

There are semi-relativistic and relativistic
methods of the FW transformation. We use the term
``semi-relativistic'' for methods \cite{FW,Steph,Reuse,ultrafast}
using an expansion of a derived block-diagonal Hamiltonian in even
terms of ascending order in $1/c$. For the semi-relativistic and the relativistic methods, the zeroth order Hamiltonian is
the Schr\"{o}dinger one and the FW Hamiltonian of a free particle,
respectively. The first semi-relativistic method has been proposed
by Foldy and Wouthuysen \cite{FW}.
A FW Hamiltonian obtained by any semi-relativistic method contains a series in powers of the momentum and potential [$p/(mc)$ and $V/(mc^2)$], while relativistic methods 
give a compact relativistic expression for this Hamiltonian for
any particle momentum. The first relativistic FW transformation
method has been presented in Ref. \cite{Morpurgo}. Some of the
relativistic methods developed in physics are based on unitary
transformations \cite{JMP,FizElem,PRA,PRA2015,PengReiher} and a number of
these methods uses different approaches \cite{relativistic}.

All FW transformation methods applied in
quantum chemistry are relativistic. The methods based on unitary
transformations follow the approach elaborated by Douglas, Kroll,
and Hess \cite{DouglasKroll,Hess} but often use different
transformation operators
\cite{QuantChem,ReiherWolf,ReiherWolfNext,local}. They allow one
to fulfill not only high-order \cite{Highorder} but also
arbitrary-order
\cite{ReiherWolf,ReiherWolfNext,arbitraryorder,arbitraryorderPH}
Douglas-Kroll-Hess (DKH) transformations. The DKH transformations
expand the FW Hamiltonian in a series in powers of the parameter
$V/\sqrt{m^2c^4+c^2\bm p^2}$ which is the potential divided by the
total kinetic energy. We can also mention the infinite-order
two-component method of Barysz and collaborators \cite{BSS} which
is the two-step exact-decoupling approach. Another successful
relativistic two-component method is the zeroth-order regular
approximation \cite{ZORA}. The exact FW transformation can also be
performed in one step \cite{X2C}. We can refer to the reviews
\cite{PengReiher,Autschbach,ReiherTCA,Liu,local,NakajimaH,ReiherArXivBook,ReiherRev}
and to the books \cite{Dyall,ReiherWolfBook} for more details.

Many transformation methods allowing one to derive a block-diagonal
Hamiltonian do not lead to the FW representation
(see Refs. \cite{E,erik,JMPcond}). It has been proven in Refs. \cite{E,erik} that
the \emph{resulting} exponential operator of the FW transformation
should be odd and Hermitian. Paradoxically, the original FW method
\cite{FW} explained in Sec. \ref{nonrlFW} does not satisfy this
requirement and does not lead to the FW representation
\cite{erik,dVFor}. Main distinctive features of the FW
transformation are considered in Sec. \ref{Eerik}. To perform the
FW transformation, one can use either the exact Eriksen method
\cite{erik} or one of two different approaches based on successive
approximations. These possibilities are analyzed in Sec.
\ref{Approaches}. The correction of the original FW method is presented in Sec. \ref{CorrHam}. We
demonstrate that the \emph{corrected} original FW method leads to
a clear and straightforward calculation of the FW Hamiltonian and
can be rather convenient for practical use. To confirm this
statement, two examples are given in Sec. \ref{Examples}.
For any correct semi-relativistic method, the
series of relativistic corrections as a whole may define the
\emph{exact} FW Hamiltonian. However, this takes place only when
this series converges. When $p/(mc)>1$ (this situation takes place
for an electron near a nucleus), the semi-relativistic methods
become inapplicable \cite{ReiherWolf}. Restrictions caused by this
circumstance are considered in Sec. \ref{Discussion} where the
results presented are discussed and summarized.


\section{Original Foldy-Wouthuysen method} \label{nonrlFW}

In the general case, a transformation to a new representation
described by the wave function $\Psi'$ is performed with the
unitary operator $U$:
\begin{equation} \Psi'=U\Psi=\exp{(iS)}\Psi. \label{eqUiS} \end{equation}

This transformation involves not only the Hamiltonian operator but also the $-i\frac{\partial}{\partial
t}$ one. As a result, the Hamiltonian operator in the new representation takes the form
\begin{equation} {\cal H}'=U\left({\cal H}-i\hbar\frac{\partial}{\partial
t}\right)U^{-1}+ i\hbar\frac{\partial}{\partial t}  \label{taeq2} \end{equation} or
\begin{equation} {\cal H}'=U{\cal H}U^{-1}-i\hbar U\frac{
\partial U^{-1}}{\partial t}. \label{eq2}
\end{equation}

The initial Hamiltonian operator 
can be split into even and odd operators commuting and anticommuting with the operator
$\beta$, respectively:
\begin{equation} {\cal H}=\beta{\cal M}+{\cal E}+{\cal
O},~~~\beta{\cal M}={\cal M}\beta, ~~~\beta{\cal E}={\cal E}\beta,
~~~\beta{\cal O}=-{\cal O}\beta. \label{eq3} \end{equation} The
even operators ${\cal M}$ and ${\cal E}$ and the odd operator
${\cal O}$ are diagonal and off-diagonal in two spinors,
respectively. This equation is applicable for a particle with any
spin if the number of components of a corresponding wave function
is equal to $2(2s+1)$, where $s$ is the spin quantum number. For a
Dirac particle, the ${\cal M}$ operator usually reduces to the
particle rest energy $mc^2$:
\begin{equation} {\cal H}_D=\beta mc^2+{\cal E}+{\cal
O}. \label{eq3Dirac} \end{equation}

The Hamiltonian ${\cal H}$ is Hermitian for fermions and pseudo-Hermitian (more exactly,
$\beta$-pseudo-Hermitian, ${\cal H}={\cal H}^\ddag\equiv\beta{\cal H}^\dag\beta$) for bosons. We assume that the
operators $\beta{\cal M},~{\cal E}$, and ${\cal O}$ also possess this property. The transformation operator for bosons is therefore $\beta$-pseudounitary
($U^\dag=\beta U^{-1}\beta$). We can mention the existence of bosonic symmetries of the Dirac equation \cite{Simulik}.

Equation (\ref{taeq2}) can be written in the form
\begin{equation} {\cal H}'-i\hbar\frac{\partial}{\partial
t}=U\left({\cal H}-i\hbar\frac{\partial}{\partial
t}\right)U^{-1}=U\left(\beta{\cal M}+{\cal E}+{\cal
O}-i\hbar\frac{\partial}{\partial
t}\right)U^{-1}.
\label{taeq3}
\end{equation}
This equation allows us to state a rather important property of the FW transformation for a particle in nonstationary (time-dependent) fields. 
Transformations of two even operators, ${\cal E}$ and
$-i\hbar\frac{\partial}{\partial t}$, are very similar. As a
result, the FW Hamiltonian (except for terms without commutators)
contains these operators only in the combination ${\cal F}={\cal
E}-i\hbar\frac{\partial}{\partial t}$. Therefore, a transition
from a stationary to a nonstationary case can be performed with a
replacement of ${\cal E}$ with ${\cal F}$ in all terms containing
commutators.

In the present work, we focus our attention on Dirac fermions.

The original FW method \cite{FW} belongs to
iteration (step-by-step) methods because a block-diagonalization
of the Hamiltonian is a result of successive iterations.
This method allows one to obtain a series of
corrections in powers of the momentum and potential to a
nonrelativistic Hamiltonian. For the Dirac Hamiltonian written in
the form (\ref{eq3Dirac}), one deduces a series in powers of
${\cal O}/(mc^2)$ and ${\cal E}/(mc^2)$ including mixed terms.
The first step of the FW transformation is performed with the
exponential operator \cite{FW}
\begin{equation}  S=-\frac{i}{2mc^2}\beta{\cal O}. \label{eq5} \end{equation}

The transformed Hamiltonian can be written in the form
\begin{equation}  \begin{array}{c}
{\cal H}'={\cal H}+i[S,{\cal H}]+\frac{i^2}{2!}[S,[S,{\cal
H}]]+\frac{i^3}{3!} [S,[S,[S,{\cal H}]]]+\dots\\-
\hbar\dot{S}-\frac{i\hbar}{2!}[S,\dot{S}]-\frac{i^2\hbar}{3!}
[S,[S,\dot{S}]]-\dots, \end{array} \label{eqNEX} \end{equation}
where $[\dots,\dots]$ means a commutator. As a result of this
transformation, we find
\begin{equation} {\cal H}'=\beta mc^2+{\cal E}'+{\cal
O}',~~~\beta{\cal E}'={\cal E}'\beta, ~~~\beta{\cal O}'=-{\cal
O}'\beta. \label{eq7}\end{equation} 
The odd operator ${\cal O}'$ is now $ O(1/m)$. The next step has been
made with the operator
\begin{equation}  S'=-\frac{i}{2mc^2}\beta{\cal O}'.
\label{eq5nt} \end{equation}
This procedure can be repeated to decrease an order of magnitude
of odd terms. It is important that the expression of ${\cal H}''$ in terms of ${\cal E}',{\cal O}'$ is the same as that of ${\cal H}'$ in terms of ${\cal E},{\cal O}$.
This property remains valid at every step.

The original FW method can be simplified if one transforms the operator ${\cal F}$ instead of performing separate transformations of the operators ${\cal E}$ and $-i\frac{\partial}{\partial
t}$.

After the third step, 
the resulting FW transformation operator is given by
\begin{equation}U_{FW}=\exp{(iS'')}\exp{(iS')}\exp{(iS)}.\label{Vvetotr} \end{equation}


The initial Dirac Hamiltonian describing a particle in an
electromagnetic field is defined by
\begin{equation}
{\cal H}_{D}=\beta mc^2+c\bm \alpha\cdot\bm \pi +e\Phi,
\label{eqIntif}\end{equation} where $\bm\pi= \bm p-\frac{e}{c}\bm
A$ is the kinetic momentum operator, $\Phi$ and $\bm A$ are scalar and vector potentials of the
electromagnetic field, respectively, and $e$ is the charge of a particle. For the electron, it is negative ($e=-|e|$).

The FW transformation results in \cite{FW}
\begin{equation}\begin{array}{c} {\cal H}_{FW}=\beta\left(mc^2+
\frac {\bm \pi^2}{2m}-\frac{\bm \pi^4}{8m^3c^2}\right)+ e\Phi-\frac{
   e\hbar}{2mc}\bm\Pi\cdot\bm B \\
   +\frac{e\hbar}{8m^2c^2}\left(\bm\Sigma\cdot[\bm\pi\times\bm
E]-\bm\Sigma
\cdot[\bm E\times\bm\pi]-\hbar\nabla\cdot\bm E\right).
\end{array} \label{3eq32} \end{equation}

So, the seminal work by Foldy and Wouthuysen has explained spin --
magnetic field coupling previously presented by the Pauli
equation, the contact (Darwin) interaction proportional to
$\nabla\cdot\bm E$, and the ``Thomas half'' in the terms describing
a spin interaction with the electric field.


Evidently, the original FW method is useless when 
the series of relativistic corrections becomes divergent. 

We can emphasize a simple cyclic form of the original FW transformation. All successive iterations are based on the general formulas (\ref{eq5}) and (\ref{eqNEX}).

\section{Main distinctive features of the Foldy-Wouthuysen transformation}\label{Eerik}

The main distinctive features of the FW transformation have been stated in Refs. \cite{E,erik}.
The even form of the final Hamiltonian was the only condition of the transformation used by Foldy and Wouthuysen.
It can be easily shown that this condition does not define the FW Hamiltonian unambiguously. The result of successive
iterations expressed by the equation
\begin{equation} U=\ldots \exp{(iS^{(n)})}\ldots \exp{(iS''')}\exp{(iS'')}\exp{(iS')}\exp{(iS)} \label{Vvetgen} \end{equation}
can be presented in the exponential form:
\begin{equation} U=\exp{(i\mathfrak{S})}.\label{Vvetott} \end{equation}
The FW Hamiltonian obtained with this operator is even. We can
perform one more unitary transformation with the operator
$U'=\exp{(i\mathfrak{T})}$, where the exponential operator
$\mathfrak{T}$ is even. This transformation does not add odd terms
to the FW Hamiltonian ${\cal H}_{FW}$. As a result, the total
transformation operator $U=U'U_{FW}$ also transforms the initial
Hamiltonian to the even form. Since $U'$ is an arbitrary even
Hermitian and unitary ($\beta$-pseudo-Hermitian and
$\beta$-pseudounitary for bosons) operator, there is an infinity
set of transformations leading the initial Hamiltonian to a
block-diagonal form.

The condition eliminating this ambiguity has been proposed by
Eriksen \cite{E} and has been substantiated by Eriksen and Kolsrud
\cite{erik}. The transformation remains \emph{unique} if the
operator $\mathfrak{S}$ in Eq. (\ref{Vvetott}) is \emph{odd},
\begin{equation} \beta\mathfrak{S}=-\mathfrak{S}\beta,
\label{VveEfrt} \end{equation} and Hermitian
($\beta$-pseudo-Hermitian for bosons). Our explanation of the
Eriksen method which is given below differs from that presented in original works
\cite{E,erik}.

Expansion of the exponential
operator (\ref{Vvetott}) in series
\begin{equation} \exp{(i\mathfrak{S})}=1+i\mathfrak{S}+\frac{(i\mathfrak{S})^2}{2!}+\frac{(i\mathfrak{S})^3}{3!}+\ldots+\frac{(i\mathfrak{S})^n}{n!}+o\bigl((i\mathfrak{S})^n\bigr) \label{exexope} \end{equation}
shows that this condition is equivalent to \cite{E,erik}
\begin{equation} \beta U_{FW}=U^\dag_{FW}\beta.\label{Erikcon} \end{equation}

We can ascertain that the arbitrary-order DKH transformation \cite{ReiherWolf,ReiherWolfNext} satisfies the condition (\ref{Erikcon}).

Thus, the FW transformation operator should satisfy Eq.
(\ref{Erikcon}) and should perform the transformation in one step.
Eriksen \cite{E} has found an operator possessing these
properties. To determine its explicit form, one can introduce the sign
operator $\lambda={\cal H}/({\cal H}^2)^{1/2}$ and can use the fact
that the operator $1+\beta\lambda$ cancels either lower or upper
spinor for positive and negative energy states, respectively. It
is easy to see that \cite{E}
\begin{equation}\lambda^2=1, ~~~ [\beta\lambda,\lambda\beta]=0, ~~~[\beta,(\beta\lambda+\lambda\beta)]=0.\label{eq3X3}
\end{equation}
Therefore, the operator of the exact FW transformation has the form
\begin{equation}
U_{E}=U_{FW}=\frac{1+\beta\lambda}{\sqrt{2+\beta\lambda+\lambda\beta}},
~~~ \lambda=\frac{{\cal H}}{({\cal H}^2)^{1/2}}. \label{eqXXI}
\end{equation}
The initial Hamiltonian operator, ${\cal H}$, is arbitrary. The even operator $\beta\lambda+\lambda\beta$ acting on the wave function with a single nonzero
spinor cannot make another spinor be nonzero.

The equivalent form of the operator $U_{E}$ \cite{JMPcond} shows that it is properly unitary ($\beta$-pseudounitary for bosons):
\begin{equation}
U_{E}=\frac{1+\beta\lambda}{\sqrt{(1+\beta\lambda)^\dag(1+\beta\lambda)}}.
\label{JMP2009}
\end{equation}

The Eriksen operator (\ref{eqXXI}) can be used for a particle with
any spin. In this case, the initial Hamiltonian is given by Eq.
(\ref{eq3}).

Evidently, the Eriksen method gives the right relativistic FW Hamiltonian for a free Dirac particle. In this case
$${\cal E}=0, ~~~ \lambda=\frac{\beta mc^2+{\cal O}}{\epsilon}, ~~~ {\cal O}=c\bm \alpha\cdot\bm p, ~~~
\sqrt{2+\beta\lambda+\lambda\beta}=2\left(1+\frac{ mc^2}{\epsilon}\right) $$ with $\epsilon=\sqrt{m^2c^4+{\cal O}^2}$.
The resulting FW Hamiltonian reads
\begin{equation}
{\cal H}_{FW}=\beta\epsilon
 \label{eqX}
\end{equation}
and coincides with the Hamiltonian derived by Foldy and Wouthuysen
\cite{FW}. A substantiation of the Eriksen method in the more
general case of ${\cal E}\neq0$ and $[{\cal O},{\cal E}]=0$  has
been fulfilled in Ref. \cite{Valid}.

Evidently, the Eriksen operator satisfies Eq. (\ref{Erikcon}). Any
additional unitary transformation violates this accordance. For
the transformation operator $U=U'U_{E}$, where
$U'=\exp{(i\mathfrak{T})}$ and the operator $\mathfrak{T}$ is
even,
block-diagonal Hamiltonians are connected by the \emph{even} exponential
transformation operator.
$$U^\dag\beta=\beta U_{E}(U')^\dag\neq\beta U'U_{E}.$$
An extra transformation with the odd operator $\mathfrak{T}$ is
also inadmissible because its action on the even Hamiltonian
${\cal H}_{FW}$ leads to an appearance of odd terms. Therefore, two
block-diagonal Hamiltonians are connected by the \emph{even} exponential
transformation operator.

\section{Approaches to the Foldy-Wouthuysen transformation}\label{Approaches}

The transformation to the FW representation can be carried out by
a lot of different methods. The Eriksen method \cite{E} is not
iterative and allows one to perform the direct FW transformation.
Equations (\ref{eqXXI}) and (\ref{JMP2009}) are convenient to
express the FW Hamiltonian as a series of relativistic corrections
on powers of ${\cal O}/m$ and ${\cal E}/m$. One can use the
formula
\begin{equation}\sqrt{{\cal H}^2}=\beta mc^2\sqrt{1+\frac{{\cal H}^2-m^2c^4}{m^2c^4}}=
\beta mc^2\sqrt{1+\frac{2\beta mc^2{\cal E}+{\cal O}^2+{\cal E}^2+
\{{\cal O},{\cal E}\}}{m^2c^4}} \label{forkorn}
\end{equation}
and expand $({\cal H}^2)^{-1/2}$ in a Taylor series \cite{E,VJ}. While
needed calculations are cumbersome, they can be made analytically
with a computer \cite{VJ}. We present below the exact FW Hamiltonian
calculated by de Vries and Jonker up to terms of the order of
$(v/c)^8$. They supposed that ${\cal O}/(mc^2)\sim(v/c)$ and ${\cal
E}/(mc^2)\sim(v/c)^2$. The result of calculations \cite{VJ} can be
presented in a more convenient form \cite{TMPFW} via multiple
commutators. While the Eriksen method is unapplicable in the
nonstationary case, we can use the property formulated below Eq. (\ref{taeq3}) and rewrite the equation obtained in Ref.
\cite{TMPFW} as follows:
\begin{equation}\begin{array}{c}
{\cal H}_{FW}=\beta\left(mc^2+ \frac {{\cal
O}^2}{2mc^2}-\frac{{\cal O}^4}{8m^3c^6}+\frac{{\cal
O}^6}{16m^5c^{10}}-\frac{5{\cal O}^8}{128m^7c^{14}}\right)\\+{\cal
E}-\frac{1}{128m^6c^{12}}\left\{(8m^4c^8-6m^2c^4{\cal O}^2+5{\cal O}^4),[{\cal
O},[{\cal O},{\cal
F}]]\right\}\\+\frac{1}{512m^6c^{12}}\left\{(2m^2c^4-{\cal O}^2),[{\cal
O}^2,[{\cal O}^2,{\cal F}]]\right\}\\
+\frac{1}{16m^3c^6}\beta\left\{{\cal O},\left[[{\cal O},{\cal
F}],{\cal F}\right]\right\}-\frac{1}{32m^4c^8}\left[{\cal
O},\left[[[{\cal O},{\cal F}],{\cal F}],{\cal
F}\right]\right]\\+\frac{11}{1024m^6c^{12}}\left[{\cal O}^2,\left[{\cal
O}^2,[{\cal O},[{\cal O},{\cal F}]]\right]\right]+A_{24},
\end{array} \label{eq12erk}
\end{equation}
where 
\begin{equation}\begin{array}{c} A_{24}=\frac{1}{256m^5c^{10}}\beta\Biggl(24\left\{{\cal O}^2, ([{\cal O},{\cal F}])^2\right\}-
20\left ([{\cal O}^2,{\cal F}]\right )^2-14\left\{{\cal O}^2,
\left [[{\cal O}^2,{\cal F}],{\cal F}\right]
\right\}\\-4\left[{\cal O},\left[{\cal O},\left[[{\cal O}^2,{\cal
F}],{\cal F}\right] \right] \right]+\frac92 \left[\left[{\cal
O},\left[{\cal O},\left[{\cal O}^2,{\cal F}\right]\right]
\right],{\cal F}\right]\\-\frac92 \left[\left[{\cal O},\left[{\cal
O},{\cal F}\right]\right],\left[{\cal O}^2,{\cal F}\right]\right]
+\frac52 \left [{\cal O}^2,\left[{\cal O},\left[[{\cal O},{\cal
F}],{\cal F}\right] \right] \right]\Biggr).
\end{array} \label{eq12A}
\end{equation}
In $A_{24}$, the first and second subscripts indicate the
respective numbers of ${\cal F}$ and ${\cal O}$ operators in the
product. A mistake in the calculation of this term made in Ref.
\cite{TMPFW} has been corrected in Ref. \cite{PRA}.

Terms of higher orders up to $(v/c)^{12}$ have also been
calculated many years ago (see Ref. \cite{VJ} and references
therein). However, the Eriksen method is not practically used in
specific calculations. Since the exact
equations (\ref{eq3X3}) and (\ref{eqXXI}) contain the square roots
of Dirac matrices, they exclude a possibility to obtain a series
of relativistic terms with the relativistic FW Hamiltonian of a
free particle \cite{FW} as the zero-order approximation. This
possibility can be realized with the relativistic methods
mentioned in Sec. \ref{Introduction}. Moreover, Eqs. (\ref{eq3X3})
and (\ref{eqXXI}) do not seem to be convenient even for deriving a
semi-relativistic FW Hamiltonian. While this possibility has been
realized (see Ref. \cite{VJ} and references therein), a necessity
to ensure a commutativity of the numerators and denominators in
the expression for $\lambda$ and in Eq. (\ref{eqXXI}) hinders a
practical application of the Eriksen method. The FW method \cite{FW} and some other semi-relativistic methods are
more straightforward.
We can note that the use of all semi-relativistic methods in
atomic physics and quantum chemistry meet some difficulties caused
by their inapplicability at $p/(mc)>1$ \cite{ReiherWolf} (see Sec.
\ref{Discussion} for more details).

Nevertheless, the calculation of the FW
Hamiltonian by the Eriksen method as a series of relativistic corrections to the zero approximation (Schr\"{o}dinger Hamiltonian) is very
important for checking results obtained by other semi-relativistic and relativistic methods.

The mostly applied approach is characterized by subsequent
iterations allowing a determination of both the exponential
operator of the FW transformation and the FW Hamiltonian.
Intermediate exponential operators obtained at each iteration
satisfy the Eriksen condition and are odd and Hermitian. This
approach can be realized by many methods
\cite{QuantChem,ReiherWolf,ReiherWolfNext,Dyall,ReiherWolfBook,Reuse,ultrafast}.
Another approach also brings the initial Hamiltonian to a
block-diagonal form after subsequent iterations but a resulting
exponential transformation operator is not odd. Since all
subsequent transformation operators are known, the resulting
exponential transformation operator can be corrected by an
elimination of even terms in the exponent. The elimination can be
made with the use of the Baker-Campbell-Hausdorff (BCH) formula
\cite{BakerCampbellHausdorff}. This formula
\cite{BakerCampbellHausdorff} defines the product of two
exponential operators:
\begin{eqnarray}
\exp(A)\exp(B)=\exp{\biggl(A+B+\frac12[A,B]+\frac{1}{12}[A,[A,B]]-\frac{1}{12}[B,[A,B]]}\nonumber\\ -\frac{1}{24}\bigl[A,[B,[A,B]]\bigr]+{\rm higher~ order~
commutators}\biggr).\label{EriKorl}\end{eqnarray}
The product of two exponential operators can be calculated with any needed accuracy \cite{Dynkin}.

When $A=iS,~B=iS'$ and the operators $S$ and $S'$ are odd and
Hermitian, the commutators $[A,B]$ and $\bigl[A,[B,[A,B]]\bigr]$
are even. Therefore, the resulting transformation operator defined
by Eq. (\ref{Vvetgen}) can be presented in the form
$U=\exp{(i\mathfrak{R})}$ where the operator $\mathfrak{R}$ is not
odd and does not satisfy the Eriksen condition (\ref{VveEfrt}).
Expansion of the operator $U$ in a power series [see Eq.
(\ref{exexope})] shows that the equivalent Eriksen condition
(\ref{Erikcon}) is also violated.

Thus, the original FW method and other iteration methods do not lead to the FW representation and give only approximate FW Hamiltonians.
This fact has been proven by Eriksen and Kolsrud \cite{erik} and later by Neznamov \cite{FizElem,VANT}.

However, there exist possibilities to correct these methods. One can perform one more unitary transformation with the transformation operator $U_{corr}$ satisfying the relation
\begin{equation}U_{corr}U=U_E,\label{Vvegtble} \end{equation}
where $U=\exp{(i\mathfrak{R})}$ is the resulting transformation operator before the correction and $U_E$ is the Eriksen operator (\ref{eqXXI}).  The solution of Eq. (\ref{Vvegtble}) reduces to an elimination of the even part of the operator $\mathfrak{R}$.
In the next section, we will consider this problem in more detail with respect to the original FW method.

\section{Correction of Hamiltonians obtained by the Foldy-Wouthuysen method}\label{CorrHam}

In the classical FW method, any subsequent exponential operator is
of a smaller order of magnitude than a preceding one. Therefore,
the BCH formula allows one to determine and eliminate an error given
by this method. Such a possibility has been first noticed by
Eriksen and Kolsrud \cite{erik}. If one can neglect $[S,[S,S']]$,
$[S',[S,S']]$, and commutators of higher orders as compared with
$[S,S']$, Eq. (\ref{EriKorl}) brings the approximate relation
\begin{eqnarray}
\exp(iS')\exp(iS)=\exp{\left(\frac12[S,S']\right)}\exp{[i(S'+S)]}.\label{EriKorlnew}\end{eqnarray}
Since the operator $i(S'+S)$ is odd and the corrected Hamiltonian is even (with a needed accuracy), the left multiplication of the FW transformation operator by the even operator
\begin{eqnarray}
U_{corr}=\exp{\left(-\frac12[S,S']\right)}
\label{FWcorro}\end{eqnarray} does not add any odd terms to the
Hamiltonian and allows one to cancel the error of the FW method in the
leading order. We can obtain from Eqs. (\ref{eq5}) and
(\ref{eq5nt}) that the commutator of the two first exponential
operators is approximately equal to \cite{dVFor}
\begin{eqnarray} [S,S']=-\frac{\beta}{8m^3c^6}[{\cal O}^2,{\cal F}].\label{DeVrs}\end{eqnarray} Therefore,
\begin{equation}
U_{corr}=\exp{\left(\frac{\beta}{16m^3c^6}[{\cal O}^2,{\cal F}]\right)}. \label{Vvetg} \end{equation}

This additional transformation eliminates the difference between
the results obtained by the original FW method \cite{FW} and the
Eriksen one \cite{E} in the leading order. For the stationary
case, it has been shown in Ref. \cite{erik}.

In the general case, the corrected transformation operator has the
form
\begin{equation}U_E=U_{corr}U\equiv U_{corr}\ldots \exp{(iS^{(n)})}\ldots \exp{(iS''')}\exp{(iS'')}\exp{(iS')}\exp{(iS)}\label{Vvetggn} \end{equation}
and it must be equal to the Eriksen operator. Since the
transformation with the operator $U$ results in the FW Hamiltonian
obtained by the original method \cite{FW}, ${\cal
H}_{FW}^{(orig)}$, the corrected (right) FW Hamiltonian is given
by
\begin{equation}
{\cal H}_{FW}=U_{corr}\left({\cal H}_{FW}^{(orig)}-i\hbar\frac{\partial}{\partial
t}\right)U_{corr}^{-1}+ i\hbar\frac{\partial}{\partial t}. \label{corrFWH} \end{equation}
Since the operators ${\cal H}_{FW}$ and ${\cal H}_{FW}^{(orig)}$ are even, the operator $U_{corr}$ is also even.

\section{Examples of the application of the corrected original method}\label{Examples}

In this section, we will consider two examples of the application
of the corrected original FW method and will
show the importance of the corrections made.

\subsection{Foldy-Wouthuysen transformation with a calculation of all terms up to the order of $(v/c)^6$}

Let us derive the FW Hamiltonian and calculate all terms up to the
order of $(v/c)^6$ on condition that ${\cal
E}/(mc^2)\sim(v/c)^2,~{\cal O}/(mc^2)\sim v/c$. The successive
steps are given by
\begin{equation}\begin{array}{c}
S=-\frac{i}{2mc^2}\beta{\cal O},~~~
{\cal H}'=\beta mc^2+ {\cal E}+\beta\left(\frac{{\cal O}^2}{2mc^2}-\frac{{\cal
O}^4}{8m^3c^6}+\frac{{\cal
O}^6}{144m^5c^{10}}\right)\\-\frac{1}{8m^2c^4}[{\cal O},[{\cal O},
{\cal F}]]+\frac{1}{384m^4c^8}\bigl[{\cal O},[{\cal O},[{\cal O},[{\cal O},
{\cal F}]]]\bigr]\\+\frac{\beta}{2mc^2}[{\cal O},{\cal F}]-\frac{{\cal
O}^3}{3m^2c^4}+\frac{{\cal
O}^5}{30m^4c^8}-\frac{\beta}{48m^3c^6}\bigl[{\cal O},[{\cal O},[{\cal O},
{\cal F}]]\bigr],\\
S'=-\frac{i}{4m^2c^4}[{\cal O},{\cal F}]+i\beta\left(\frac{{\cal O}^3}{6m^3c^6}-\frac{{\cal
O}^5}{60m^5c^{10}}\right)+\frac{i}{96m^4c^8}\bigl[{\cal O},[{\cal O},[{\cal O},
{\cal F}]]\bigr],\\
{\cal H}''=\beta mc^2+ {\cal E}+\beta\left(\frac{{\cal O}^2}{2mc^2}-\frac{{\cal
O}^4}{8m^3c^6}+\frac{{\cal
O}^6}{16m^5c^{10}}\right)-\frac{1}{8m^2c^4}[{\cal O},[{\cal O},
{\cal F}]]\\-\frac{\beta}{8m^3c^6}\left([{\cal O},
{\cal F}]\right)^2+\frac{3}{64m^4c^8}\bigl\{{\cal O}^2,[{\cal O},[{\cal O},
{\cal F}]]\bigr\}+\frac{5}{128m^4c^8}[{\cal O}^2,[{\cal O}^2,
{\cal F}]]\\+\frac{1}{4m^2c^4}[[{\cal O},
{\cal F}],{\cal F}]-\frac{\beta}{6m^3c^6}[{\cal O}^3,
{\cal F}]-\frac{\beta}{8m^3c^6}\{{\cal O}^2,[{\cal O},
{\cal F}]\},\\
S''=-\frac{i\beta}{8m^3c^6}[[{\cal O},
{\cal F}],{\cal F}]+\frac{i}{12m^4c^8}[{\cal O}^3,
{\cal F}]+\frac{i}{16m^4c^8}\{{\cal O}^2,[{\cal O},
{\cal F}]\}.
\end{array} \label{Proeq}\end{equation}

After the transformation with the operator $S''$, the final Hamiltonian obtained by the original FW method takes the form
\begin{equation}\begin{array}{c}
{\cal H}_{FW}^{(orig)}=\beta mc^2+ {\cal E}+\beta\left(\frac{{\cal O}^2}{2mc^2}-\frac{{\cal
O}^4}{8m^3c^6}+\frac{{\cal
O}^6}{16m^5c^{10}}\right)-\frac{1}{8m^2c^4}[{\cal O},[{\cal O},
{\cal F}]]\\-\frac{\beta}{8m^3c^6}\left([{\cal O},
{\cal F}]\right)^2+\frac{3}{64m^4c^8}\bigl\{{\cal O}^2,[{\cal O},[{\cal O},
{\cal F}]]\bigr\}+\frac{5}{128m^4c^8}[{\cal O}^2,[{\cal O}^2,
{\cal F}]].
\end{array} \label{ProF1}\end{equation}

It should be corrected according to Eq. (\ref{corrFWH}). In this case, the needed accuracy can be achieved with the single commutator:
\begin{equation}\begin{array}{c}
{\cal H}_{FW}={\cal H}_{FW}^{(orig)}-\Biggl[\frac12[S,S'],\left({\cal H}_{FW}^{(orig)}-i\hbar\frac{\partial}{\partial
t}\right)\Biggr].
\end{array} \label{ProeR}\end{equation}
With the use of Eqs. (\ref{Proeq}) and (\ref{ProF1}), we obtain
\begin{equation}\begin{array}{c}
{\cal H}_{FW}={\cal H}_{FW}^{(orig)}+\Biggl[\frac{\beta}{16m^3c^6}[{\cal O}^2,{\cal F}],\left({\cal F}+\beta\frac{{\cal O}^2}{2mc^2}\right)\Biggr].
\end{array} \label{Proeo}\end{equation}

As a result,
\begin{equation}\begin{array}{c}
{\cal H}_{FW}=\beta mc^2+ {\cal E}+\beta\left(\frac{{\cal O}^2}{2mc^2}-\frac{{\cal
O}^4}{8m^3c^6}+\frac{{\cal
O}^6}{16m^5c^{10}}\right)-\frac{1}{8m^2c^4}[{\cal O},[{\cal O},
{\cal F}]]\\+\frac{\beta}{16m^3c^6}\left\{{\cal O},\left[[{\cal O},{\cal
F}],{\cal F}\right]\right\}+\frac{3}{64m^4c^8}\left\{
{\cal O}^2,[{\cal O},[{\cal O},{\cal
F}]]\right\}+\frac{1}{128m^4c^8}[{\cal O}^2,[{\cal O}^2,
{\cal F}]].
\end{array} \label{ProE1}\end{equation}
This equation agrees with the result obtained by the Eriksen method and expressed by Eq. (\ref{eq12erk}).

\subsection{Foldy-Wouthuysen transformation with a calculation of all terms up to the order of $m^{-4}$}

It is also instructive to calculate all terms up to the order of
$m^{-4}$ on the condition that ${\cal E}\sim{\cal O}$. In this case,
the successive exponential transformation operators are given by
\begin{equation}\begin{array}{c}
S=-\frac{i}{2mc^2}\beta{\cal O},
\\
S'=-\frac{i}{4m^2c^4}[{\cal O},{\cal F}]+\frac{i}{6m^3c^6}\beta{\cal O}^3+\frac{i}{96m^4c^8}\bigl[{\cal O},[{\cal O},[{\cal O},
{\cal F}]]\bigr],\\
S''=-\frac{i\beta}{8m^3c^6}[[{\cal O},
{\cal F}],{\cal F}]+\frac{i}{12m^4c^8}[{\cal O}^3,
{\cal F}]+\frac{i}{16m^4c^8}\{{\cal O}^2,[{\cal O},
{\cal F}]\},\\
S'''=-\frac{i}{16m^4c^8}\bigl[[[{\cal O},
{\cal F}],{\cal F}],{\cal F}\bigr].
\end{array} \label{Pro2p}\end{equation}
The Hamiltonian has the form
\begin{equation}\begin{array}{c}
{\cal H}_{FW}^{(orig)}=\beta mc^2+ {\cal E}+\beta\left(\frac{{\cal O}^2}{2mc^2}-\frac{{\cal
O}^4}{8m^3c^6}\right)-\frac{1}{8m^2c^4}[{\cal O},[{\cal O},
{\cal F}]]-\frac{\beta}{8m^3c^6}\left([{\cal O},
{\cal F}]\right)^2\\+\frac{3}{64m^4c^8}\bigl\{{\cal O}^2,[{\cal O},[{\cal O},
{\cal F}]]\bigr\}+\frac{5}{128m^4c^8}[{\cal O}^2,[{\cal O}^2,
{\cal F}]]+\frac{1}{32m^4c^8}\bigl[[{\cal O},
{\cal F}],[[{\cal O},{\cal F}],{\cal F}]\bigr].
\end{array} \label{ProF2}\end{equation}

To calculate all terms up to the order of $m^{-4}$, one needs take into account commutators with the exponential transformation operators $S'$ and $S''$:
\begin{equation}\begin{array}{c}
{\cal H}_{FW}={\cal H}_{FW}^{(orig)}-\Biggl[\frac12[S,(S'+S'')],\left({\cal H}_{FW}^{(orig)}-i\hbar\frac{\partial}{\partial
t}\right)\Biggr].
\end{array} \label{ProeT}\end{equation} The implication of the operator $S''$ into the correction procedure distinguishes this example from the precedent one.

Equations (\ref{Pro2p})--(\ref{ProeT}) result in
\begin{equation}\begin{array}{c}
{\cal H}_{FW}={\cal
H}_{FW}^{(orig)}+\Biggl[\frac{\beta}{16m^3c^6}[{\cal O}^2,{\cal
F}],\left({\cal F}+\beta\frac{{\cal
O}^2}{2mc^2}\right)\Biggr]\\-\frac{1}{32m^4c^8}\Bigl[\bigl[{\cal
O},[[{\cal O},{\cal F}],{\cal F}]\bigr],{\cal F}\Bigr].
\end{array} \label{Proen}\end{equation}
Since $\bigl[[{\cal O}, {\cal F}],[[{\cal O},{\cal F}],{\cal
F}]\bigr]-\Bigl[\bigl[{\cal O},[[{\cal O},{\cal F}],{\cal
F}]\bigr],{\cal F}\Bigr]=-\Bigl[{\cal O},\bigl[[[{\cal O},{\cal
F}],{\cal F}],{\cal F}\bigr]\Bigr]$, the corrected FW Hamiltonian
has the form
\begin{equation}\begin{array}{c}
{\cal H}_{FW}=\beta mc^2+ {\cal E}+\beta\left(\frac{{\cal O}^2}{2mc^2}-\frac{{\cal
O}^4}{8m^3c^6}\right)-\frac{1}{8m^2c^4}[{\cal O},[{\cal O},
{\cal F}]]\\+\frac{\beta}{16m^3c^6}\left\{{\cal O},\left[[{\cal O},{\cal
F}],{\cal F}\right]\right\}+\frac{3}{64m^4c^8}\left\{
{\cal O}^2,[{\cal O},[{\cal O},{\cal
F}]]\right\}\\+\frac{1}{128m^4c^8}[{\cal O}^2,[{\cal O}^2,
{\cal F}]]-\frac{1}{32m^4c^8}\Bigl[{\cal O},\bigl[[[{\cal O},{\cal F}],{\cal F}],{\cal F}\bigr]\Bigr].
\end{array} \label{ProE2}\end{equation}
This expression also agrees with Eq. (\ref{eq12erk}).

We can conclude that the corrected original FW method ensures a straightforward derivation of the FW Hamiltonian.

\section{Discussion and summary}\label{Discussion}

The wonderful achievements of Eriksen are the formulation and the
substantiation of conditions of transformation to the FW
representation, the derivation of the exact FW transformation
operator, the proof of an approximateness of the original
FW method, and the discovery of the possibility of
its correction. The Eriksen method gives the correct FW
Hamiltonian for a free particle and also in the more general case
\cite{Valid} of ${\cal E}\neq0$ and $[{\cal O},{\cal E}]=0$.
In the general nonrelativistic case, the
Eriksen formula (\ref{eqXXI}) allows one to present the FW
Hamiltonian as a series of relativistic corrections to the
Schr\"{o}dinger Hamiltonian (see Ref. \cite{VJ}). However, this
series (as well as a series given by any nonrelativistic method)
is divergent when $p/(mc)>1$.

It has been shown in Ref. \cite{ReiherWolf}
that the use of all semi-relativistic methods in quantum chemistry
is restricted due to their divergence at $p/(mc)>1$. Evidently,
this takes place in a small region near a nucleus. In the
classical theory, the energy of the electron is given by
$$E=\sqrt{m^2c^4+c^2{\bm p}^2}-\frac{Ze^2}{r},$$ where $Z$ is the atomic number. The small region of the series divergence is defined by the approximate condition \begin{equation}\begin{array}{c}
r\lesssim Zr_0, \end{array} \label{appcn}\end{equation} where
$r_0=e^2/(mc^2)= 2.818\times10^{-13}$ cm is the classical electron
radius. In this small region, FW wave eigenfunctions are
undefined. When this is not admissible, one should use appropriate
relativistic methods. For example, the method developed in Ref.
\cite{ReiherWolf} gives a convergent series because the expansion
parameter contains the kinetic energy in the denominator and is
always less than 1.

Inside of the region of the series convergence,
semi-relativistic and relativistic methods of the FW
transformation should give equivalent results.  However, an
existence of the series divergence restricts an application of all
semi-relativistic methods in quantum chemistry. The corrected FW
method is perfect in all cases when the series divergence does not
appear (for example,  for a description of a particle in a trap).
%
The derivation of the Hamiltonian ${\cal H}_{FW}^{(orig)}$
reproduced in Sec. \ref{nonrlFW} represents a straightforward
computer cycle based on the general formulas (\ref{eq5}) and
(\ref{eqNEX}). The next computer cycle is the calculation of the
resulting exponential transformation operator with the BCH formula
(\ref{EriKorl}). The product of two exponential operators defined
by this formula can be obtained with any needed accuracy. Then,
one needs to find the operator $U_{corr}$ satisfying the relation
(\ref{Vvegtble}) and eliminating the even part of the exponential
operator $\mathfrak{R}$. The FW Hamiltonian can by finally
obtained by the transformation of the operator ${\cal
H}_{FW}^{(orig)}$ with the operator $U_{corr}$. This
transformation is given by Eq. (\ref{corrFWH}). An applicability
of the corrected FW method is demonstrated by the two examples
presented in Sec. \ref{Examples}.

We can conclude that the correction of iterative methods with the
BCH formula allows one to use these methods for a derivation of
the FW Hamiltonians with a needed accuracy but
their applicability is restricted by the condition of the series
convergence.

Let us also consider the electron density at the position $\bm r_A$
of a specific nucleus $A$. In Refs. \cite{ReiherWolfBook,electrondensity}, the corresponding Dirac operator has been found to be
$$ \begin{array}{c} \widehat{O}=\sum^N_i{\widehat{O}(\bm r_i)} \qquad {\rm with} \qquad \widehat{O}(\bm r_i)=\delta^{(3)}(\bm r_i-\bm r_A)\\
=\delta(x_i-x_A)\delta(y_i-y_A)\delta(z_i-z_A). \end{array}$$
In the FW representation, this operator takes the form $U_{FW}\widehat{O}U_{FW}^\dag$. The expectation value for the electron density then reads \cite{ReiherWolfBook,electrondensity}
\begin{equation} \rho_{ii}(\bm r)=\left\langle U_{FW}\psi_i|U_{FW}\widehat{O}U_{FW}^\dag|U_{FW}\psi_i\right\rangle.\label{Vvetotf} \end{equation}
This formula has been used for specific calculations \cite{electrondensity}.
The charge distribution obtained with the FW transformation can significantly differ from the corresponding nonrelativistic charge distribution.

\section*{Acknowledgements}

This work was supported in part by the Belarusian Republican Foundation for
Fundamental Research (Grant No. $\Phi$14D-007) and by the Heisenberg-Landau Program of the German Ministry for Science and Technology (Bundesministerium f\"{u}r Bildung und Forschung).


\begin{thebibliography}{99}

\bibitem{FW}
 L.\,L. Foldy, S.\,A. Wouthuysen, 
Phys. Rev. \textbf{78}, 29 (1950).

\bibitem{NW}
T.\,D. Newton, E.\,P. Wigner,
Rev. Mod. Phys. \textbf{21}, 400 (1949).

\bibitem{JMP}
A.\,J. Silenko, J. Math. Phys. {\bf 44}, 2952 (2003).

\bibitem{JINRLett12}
A.\,J. Silenko, Pis'ma Zh. Fiz. Elem. Chast. Atom. Yadra \textbf{10},
144 (2013) [Phys. Part. Nucl. Lett. \textbf{10}, 91 (2013)].

\bibitem{CMcK}
J.\,P. Costella, B.\,H.\,J. McKellar,
Am. J. Phys. \textbf{63}, 1119 (1995).

\bibitem{QuantChem}
D. K\c{e}dziera and M. Barysz, Chem. Phys. Lett. \textbf{446}, 176 (2007);
F. Aquilante \emph{et al}, J. Comput. Chem. \textbf{31}, 224 (2010);
D. Peng, N. Middendorf, F. Weigend, and M. Reiher, J. Chem. Phys. \textbf{138}, 184105 (2013).

\bibitem{ReiherWolf}
M. Reiher and A. Wolf, J. Chem. Phys. \textbf{121}, 2037 (2004).

\bibitem{ReiherWolfNext}
M. Reiher and A. Wolf, J. Chem. Phys. \textbf{121}, 10945 (2004); A. Wolf and M. Reiher, J. Chem. Phys. \textbf{124}, 064102 (2006); \textbf{124}, 064103 (2006).

\bibitem{Dyall} K.\,G. Dyall and K. Faegri, \textit{Introduction to relativistic quantum chemistry} (Oxford University Press, Oxford, 2007).

\bibitem{ReiherWolfBook}
M. Reiher and A. Wolf, \textit{Relativistic Quantum Chemistry: The Fundamental Theory of Molecular Science} (Wiley-VCH, Weinheim, 2009).

\bibitem{ReiherArXivBook}
M. Reiher, \textit{Sequential decoupling of negative-energy states
in Douglas–Kroll–Hess theory}. In: \textit{Handbook of Relativistic Quantum Chemistry}, ed. by W. Liu ((Springer-Verlag, Berlin, 2015).

\bibitem{Autschbach}
J. Autschbach, Coord. Chem. Rev. \textbf{251}, 1796 (2007).

\bibitem{NakajimaH}
T. Nakajima, K. Hirao, 
Chem. Rev. \textbf{112}, 385 (2012). 

\bibitem{electrondensity}
R. Mastalerz, R. Lindh, M. Reiher, 
Chem. Phys. Lett. \textbf{465}, 157 (2008). 

\bibitem{ReiherRev}
M. Reiher, 
WIREs Comput. Mol. Sci. \textbf{2}, 139 (2012). 

\bibitem{Steph}
S. Stephani, 
Ann. Phys. (Leipzig) {\bf 470}, 12 
(1965).

\bibitem{Reuse} F.\,A. Reuse, \textit{Electrodynamique et Optique Quantiques} (Presses Polytechniques et
Universitaires Romandes, Lausanne, 2007).

\bibitem{ultrafast}
Y. Hinschberger and P.-A. Hervieux, 
\textit{Phys. Lett. A} \textbf{376}, 813 (2012).

\bibitem{Morpurgo}
G. Morpurgo, Nuovo Cimento \textbf{15}, 624 (1960).

\bibitem{FizElem} V. P. Neznamov, Fiz. Elem. Chastits At. Yadra {\bf
37}, 152 (2006) [Phys. Part. Nucl. {\bf 37}, 86 (2006)].

\bibitem{PRA}
A.\,J. Silenko, Phys. Rev. A \textbf{77}, 012116 (2008).

\bibitem{PRA2015}
A.\,J. Silenko, Phys. Rev. A \textbf{91}, 022103 (2015).

\bibitem{relativistic}
E.\,I. Blount, Phys. Rev. \textbf{128}, 2454 (1962);
A.\,J. Silenko, Theor. Math. Phys. {\bf 105}, 1224 
(1995);
{\bf 112}, 922 (1997); 
K.\,Y. Bliokh, Europhys. Lett. \textbf{72}, 7
(2005);
Phys. Lett. A \textbf{351}, 123 (2006);
P. Gosselin, A. Berard, and H. Mohrbach, Eur. Phys. J. B \textbf{58},
137 
(2007);
Phys. Lett. A \textbf{368}, 356 
(2007);
P. Gosselin, J. Hanssen, and H. Mohrbach, Phys. Rev. D
\textbf{77}, 085008 (2008);
P. Gosselin and H. Mohrbach, Eur. Phys. J. C \textbf{64}, 495 
(2009).

\bibitem{DouglasKroll}
M. Douglas and N.\,M. Kroll, Ann. Phys. \textbf{82}, 89 (1974).

\bibitem{Hess}
B.\,A. Hess, Phys. Rev. A \textbf{32}, 756 (1985);
\textbf{33}, 3742 (1986).

\bibitem{ReiherTCA}
M. Reiher, Theor. Chem. Acc. \textbf{116}, 241 (2006).

\bibitem{Liu}
W. Liu, Mol. Phys. \textbf{108}, 1679 (2010).

\bibitem{local}
D. Peng and M. Reiher, 
J. Chem. Phys. \textbf{136}, 244108 (2012).

\bibitem{PengReiher}
D. Peng and M. Reiher, 
Theor. Chem. Acc. \textbf{131}, 1081 (2012).

\bibitem{Highorder}
T. Nakajima and K. Hirao, Chem. Phys. Lett. \textbf{329}, 511 (2000);
J. Chem. Phys. \textbf{113}, 7786 (2000);
A. Wolf, M. Reiher, and B. A. Hess, J. Chem. Phys. \textbf{117}, 9215 (2002);
C. van Wüllen, J. Chem. Phys. \textbf{120}, 7307 (2004).
\bibitem{E}
E. Eriksen, 
Phys. Rev. \textbf{111}, 1011 
(1958).

\bibitem{arbitraryorder}
M. Reiher and A. Wolf,
Phys. Lett. A \textbf{360}, 603 (2007).

\bibitem{arbitraryorderPH}
D. Peng and K. Hirao, J. Chem. Phys. \textbf{130}, 044102 (2009).

\bibitem{ZORA}
C. Chang, M. Pelissier, and P. Durand, Phys. Scr. \textbf{34}, 394 (1986);
E. van Lenthe, E. J. Baerends, and J. G. Snijders, J. Chem. Phys. \textbf{99}, 4597
(1993); \textbf{101}, 9783 (1994).

\bibitem{BSS}
M. Barysz, A. J. Sadlej, J. G. Snijders,
Int. J. Quantum Chem. \textbf{65}, 225 (1997); 
M. Barysz, J. Chem. Phys. \textbf{114}, 9315 (2001);
M. Barysz, A. J. Sadlej, 
J. Chem. Phys. \textbf{116}, 2696 (2002); 
D. K\c{e}dziera and M. Barysz, Chem. Phys. Lett. \textbf{393}, 521 (2004).

\bibitem{X2C}
K. G. Dyall, J. Chem. Phys. \textbf{106}, 9618 (1997); \textbf{109}, 4201 (1998); \textbf{115}, 9136 (2001); J. Comput. Chem. \textbf{23}, 786 (2002);
K. G. Dyall and T. Enevoldsen, J. Chem. Phys. \textbf{111}, 10000 (1999);
M. Filatov and D. Cremer, J. Chem. Phys. \textbf{119}, 11526 (2003); \textbf{122}, 064104 (2005);
W. Kutzelnigg and W. Liu, J. Chem. Phys. \textbf{123}, 241102 (2005); Mol. Phys. \textbf{104}, 2225 (2006);
W. Liu and D. Peng, J. Chem. Phys. \textbf{125}, 044102 (2006); M. Filatov and K. G. Dyall, Theor. Chem. Acc. \textbf{117}, 333 (2007); W. Liu and W. Kutzelnigg, J. Chem. Phys. \textbf{126}, 114107 (2007); M. Ilia\u{s} and T. Saue, J. Chem. Phys. \textbf{126}, 064102 (2007);
D. Peng, W. Liu, Y. Xiao, and L. Cheng, J. Chem. Phys. \textbf{127}, 104106
(2007); W. Liu and D. Peng, J. Chem. Phys. \textbf{131}, 031104 (2009); J. Sikkema, L. Visscher, T. Saue, and M. Ilia\u{s}, J. Chem. Phys. \textbf{131}, 124116
(2009).



\bibitem{erik}
E. Eriksen and M. Korlsrud, Nuovo Cimento Suppl. \textbf{18},
1 (1960).

\bibitem{JMPcond}
V.\,P. Neznamov and A.\,J. Silenko, J. Math. Phys. \textbf{50},
122302 (2009).

\bibitem{dVFor}
E. de Vries, 
Fortschr. Phys. \textbf{18}, 149 (1970).

\bibitem{Simulik} V.\,M. Simulik, I.\,Yu. Krivsky, Bosonic symmetries of the massless Dirac equation
Adv. Appl. Clifford Alg. \textbf{8}, 
69 (1998);
Phys. Lett. A \textbf{375}, 
2479 (2011);
V.\,M. Simulik, I.\,Yu. Krivsky, I.\,L. Lamer,
Ukr. J. Phys. 
\textbf{58}, 523 (2013);
TWMS Journ. Appl. Engin. Math. 
\textbf{3}, 46 (2013).

\bibitem{Valid} A.\,J. Silenko, Pis'ma Zh. Fiz. Elem. Chast. Atom. Yadra \textbf{10},
321 (2013) [Phys. Part. Nucl. Lett. \textbf{10}, 198 (2013)].

\bibitem{VJ}
E. de Vries, J.\,E. Jonker,
Nucl. Phys. B \textbf{6}, 213 
(1968).

\bibitem{TMPFW}
A.\,J. Silenko, 
Teor. Mat. Fiz. \textbf{176}, 189 (2013)
[Theor. Math. Phys. \textbf{176}, 987 
(2013)].

\bibitem{BakerCampbellHausdorff} H. Baker, Proc. Lond. Math. Soc. (1) \textbf{34}, 347 (1902); (1) \textbf{35},  333 (1903); (Ser. 2) \textbf{3}, 24 (1905);
J. Campbell, Proc. Lond. Math. Soc. \textbf{28}, 381 (1897); \textbf{29}, 14 (1898);
F. Hausdorff, Ber. Verh. Saechs. Akad. Wiss. Leipzig, Math.-Phys.
Kl. \textbf{58}, 19 (1906). This formula has also been obtained by H. Poincar\'{e}, Compt. Rend. Acad. Sci. Paris \textbf{128}, 1065 (1899); Camb. Philos. Trans. \textbf{18}, 220 (1899).

\bibitem{Dynkin}
E.\,B. Dynkin,
Dokl. Akad. Nauk SSSR \textbf{57}, 323 (1947) (in Russian).

\bibitem{VANT}  V.\,P. Neznamov, Voprosy Atomnoj Nauki i Tekhniki,
Seriya Teoreticheskaya i Prikladnaya Fizika (in Russian) {\bf 2}, 21 (1988).

\end{thebibliography}
\end{document}